\documentclass[10pt,journal]{IEEEtran}
\makeatletter
\def\ps@headings{%
\def\@oddhead{\mbox{}\scriptsize\rightmark \hfil \thepage}%
\def\@evenhead{\scriptsize\thepage \hfil \leftmark\mbox{}}%
\def\@oddfoot{}%
\def\@evenfoot{}}
\makeatother
\usepackage{amssymb}
\usepackage{amsfonts,amsmath}
\usepackage[ruled]{algorithm}
\usepackage{algpseudocode}
\usepackage{algorithmicx}
\usepackage{graphicx,epsfig,subfigure}
\usepackage{subfig}
\usepackage{color}
\usepackage[table,xcdraw]{xcolor}
\usepackage{booktabs}

\usepackage{cite}

\newcommand {\mymarginpar}[1]{\marginpar{#1}}
\renewcommand {\marginpar}[1]{}

\def\_{\rule{.3em}{.15ex}}      

\newcommand{\ls}[1]
   {\dimen0=\fontdimen6\the\font
    \lineskip=#1\dimen0
    \advance\lineskip.5\fontdimen5\the\font
    \advance\lineskip-\dimen0
    \lineskiplimit=.9\lineskip
    \baselineskip=\lineskip
    \advance\baselineskip\dimen0
    \normallineskip\lineskip
    \normallineskiplimit\lineskiplimit
    \normalbaselineskip\baselineskip
    \ignorespaces
   }

\newcommand {\bearn}{\begin{eqnarray*}}
\newcommand {\eearn}{\end{eqnarray*}}
\newcommand {\barr}{\begin{array}}
\newcommand {\earr}{\end{array}}

\newcommand {\N}{{\cal N}}

\newtheorem{definition}{Definition}
\newtheorem{property}[definition]{Property}
\newtheorem{proposition}[definition]{Proposition}
\newtheorem{lemma}[definition]{Lemma}
\newtheorem{theorem}[definition]{Theorem}
\newtheorem{corollary}[definition]{Corollary}
\newtheorem{example}[definition]{Example}
\newtheorem{remark}[definition]{Remark}

\newcommand {\benum} {\begin{enumerate}}
\newcommand {\eenum} {\end{enumerate}}

\newcommand {\bdesc} {\begin{description}}
\newcommand {\edesc} {\end{description}}

\newcommand {\bfig}[2] {\begin{figure}
  \centering
  \includegraphics[width=#2]{#1}}
\newcommand {\brotatefig}[2] {\begin{figure}[htbp]
                        \centerline {
                         \epsfig{figure={#1},clip=,angle=-90,width={#2}}}}
\newcommand {\bfigfirst}[2] {\begin{figure}[h]
                        \centerline {
                        \setlength{\epsfxsize}{#2}
                        \epsffile{#1}}}
\newcommand {\efig}[2]{ \caption{#2}
                        \label{fig:#1}
                        \end{figure}
                        \mymarginpar{fig:#1}}
\newcommand {\erotatefig}[2]{ \caption{#2}
                        \label{fig:#1}
                        \end{figure}
                        \mymarginpar{fig:#1}}
\newcommand {\rfig}[1]{Figure \ref{fig:#1}}

\newcommand {\btab}[1]{
                       \begin{table}
                       \centering
                       \begin{tabular}{#1}}
\newcommand {\etab}[3] {
                       \end{tabular}
                       \caption[#3]{#2}
                       \label{tab:#1}
                       \end{table}
                       \mymarginpar{tab:#1}
                       \vspace{.1in}}

\newcommand {\btabular}[1]{\begin{center}
                       \begin{tabular}{#1}}
\newcommand {\etabular}{\end{tabular}
                       \end{center}}

\newcommand {\bdefin}[1]{\begin{definition}
                      \mymarginpar{def:#1}
                      \label{def:#1} }
\newcommand {\edefin}       {\end{definition}}

\newcommand {\bpro}[1]{\begin{property}
                      \mymarginpar{pro:#1}
                      \label{pro:#1} }
\newcommand {\epro}   {\end{property}}

\newcommand {\bprop}[1]{\begin{proposition}
                      \mymarginpar{prop:#1}
                      \label{prop:#1} }
\newcommand {\eprop}       {\end{proposition}}

\newcommand {\blem}[1]{\begin{lemma}
                      \mymarginpar{lem:#1}
                      \label{lem:#1} }
\newcommand {\elem}   {\end{lemma}}

\newcommand {\bthe}[1]{\begin{theorem}
                      \mymarginpar{the:#1}
                      \label{the:#1} }
\newcommand {\ethe}   {\end{theorem}}
\newcommand {\rthe}[1]{Theorem \ref{the:#1}}

\newcommand {\bproof}{\noindent {\bf Proof.} \ }
\newcommand {\eproof} {\hfill \squares \\ \vspace{.3cm}}
\newcommand {\bcor}[1]{\begin{corollary}
                      \mymarginpar{cor:#1}
                      \label{cor:#1} }
\newcommand {\ecor}   {\end{corollary}}

\newcommand {\bax}[1]{\begin{axiom}
                      \mymarginpar{ax:#1}
                      \label{ax:#1} }
\newcommand {\eax}       {\vspace{-.1in} \end{axiom}}

\newcommand {\bex}[2]{\vspace{.1in}
                      \begin{example}
                      \mymarginpar{ex:#1}
                       {\bf #2}
                      \label{ex:#1} }
\newcommand {\eex}       {\end{example} \vspace{.3cm} }

\newcommand {\brem}[1]{\begin{remark}
                      \mymarginpar{rem:#1}
                      \label{rem:#1} \em }
\newcommand {\erem}   {\end{remark}}

\newcommand {\beq}[1]{\mymarginpar{eq:#1}
                      \begin{equation}
                      \label{eq:#1} }

\newcommand {\beqno}[1]{\mymarginpar{eq:#1}
                      \begin{eqnarray}
                      \nonumber}

\newcommand {\eeq}       {\end{equation}}
\newcommand {\eeqno}       { && \end{eqnarray}}
\newcommand {\req}[1]{(\ref{eq:#1})}

\newcommand {\bear}[1]{\mymarginpar{eq:#1}
                       \begin{eqnarray}
                       \label{eq:#1} }

\newcommand {\bearno}[1]{\mymarginpar{eq:#1}
                       \begin{eqnarray}
                       \nonumber}

\newcommand {\eear}{\end{eqnarray}}
\newcommand {\eearno}{\end{eqnarray}}
\newcommand {\bsel}{\left \{ \begin{array}{cl}}
\newcommand {\esel}{\end{array} \right.}

\newcommand {\bmat}[1]{\left [ \begin{array}{#1}}
\newcommand {\emat}{\end{array} \right ]}
\newcommand {\bsec}[2]{\mymarginpar{sec:#2}
                       \section{#1}
                       \label{sec:#2} }

\newcommand {\rsec}[1]{Section \ref{sec:#1}}

\newcommand {\bsubsec}[2]{\mymarginpar{sec:#2}
                       \subsection{#1}
                       \label{sec:#2} }

\def\R{I\kern-0.30em R}
\def\N{I\kern-0.30em N}
\def\P{I\kern-0.30em P}
\newcommand\squares{\vrule height6pt width7pt depth1pt}

\def\ex{{\bf\sf E}}
\def\pr{{\bf\sf P}}

\newcommand{\peri}{p}

\begin{document}

\title{Using Locality-sensitive Hashing for Rendezvous Search}

\author{Guann-Yng Jiang and Cheng-Shang Chang, ~\IEEEmembership{Fellow,~IEEE}\\
Institute of Communications Engineering\\
National Tsing Hua University \\
Hsinchu 30013, Taiwan, R.O.C. \\
Email:  kittyjiang880831@gmail.com; cschang@ee.nthu.edu.tw}

\maketitle

\begin{abstract}
The multichannel rendezvous problem is a fundamental problem for neighbor discovery in many IoT applications.
The existing works in the literature focus mostly on improving the worst-case performance, and the average-case performance is often not as good as that of the random algorithm. As IoT devices (users) are close to each other,  their available channel sets, though they might be different, are {\em similar}.
Using the locality-sensitive hashing (LSH) technique in data mining, we propose channel hopping algorithms that
  exploit the similarity between the two available channel sets to increase the rendezvous probability.
    For the synchronous setting, our algorithms have the expected time-to-rendezvous (ETTR)
  inversely proportional to a well-known similarity measure called the Jaccard index.
  For the asynchronous setting, we use dimensionality reduction to speed up the rendezvous process.
    Our numerical results  show that  our algorithms can outperform the random algorithm
    in terms of ETTR.
\end{abstract}




\bsec{Introduction}{introduction}

The multichannel rendezvous problem (MRP) that asks two users to meet each other by hopping over their available channels is a fundamental problem for neighbor discovery in many IoT applications. This problem has received much attention lately (see, e.g.,  the excellent book \cite{Book} and references therein). As classified in the paper \cite{GAP2019}, there are three key elements in the multichannel rendezvous problem: (i) users, (ii) time, and (iii) channels. Based on the assumptions on users, time, and
channels, channel hopping (CH) algorithms can be classified into various settings.
One of the most difficult settings, known as the obvious setting in the book \cite{Book}, is
the sym/async/hetero/local MRP.
In such a setting, (i) sym: users are indistinguishable, (ii)
 async: users' clocks may not be synchronized to the global clock, (iii) hetero: users may not have the same available channel sets, and (iv) local: the labels of the channels of a user are locally labeled, and they may not be the same as the global labels of the channels. In such a setting, nothing can be learned from a failed rendezvous attempt.
 It was shown in \cite{ToN2017} that the expected time-to-rendezvous (ETTR) in such a setting is lower bounded $(n_1 n_2+1)/(n_{1,2}+1)$, where $n_1$ (resp. $n_2$) is the number of available channels of user 1 (resp. user 2), and $n_{1,2}$ is the number of commonly available channels between users $1$ and $2$. The random algorithm in which each user randomly selects a channel from its available channel set performs amazing well as
 the ETTR of the random algorithm is $n_1 n_2/n_{1,2}$. Such an ETTR  is very close to the lower bound when $n_{1,2}$ is not too small. In other words, the random algorithm is nearly optimal in ETTR for the sym/async/hetero/local MRP.

 The obvious setting; however, may not be realistic for practical IoT applications due to the following two reasons: (i) the channels are usually globally labeled (according to some international standards),
 and (ii) the available channel sets of two users are similar as they are in the same proximity.
 With the first feature (assumption), the rendezvous search problem falls into the category
 of the sym/async/hetero/global MRP in \cite{ToN2017}, where the labels of the channels of a user are {\em globally} labeled.   For that setting, there are many existing
 algorithms that focus on the maximum time-to-rendezvous (MTTR). In particular, it has been shown that the MTTR of the CH sequences in \cite{Chen14,Improved2015,Chang18} is $O(\log\log N)n_1 n_2$ for an MRP with $N$ globally labeled channels. However, the ETTRs of these CH sequences are not as good as that of the random algorithm.
 Thus, one might pose the question of whether the ETTR can be further reduced by using the second feature.

 To address such a question, we apply the locality-sensitive hashing (LSH) technique that is widely used in data mining.
 To the best of our knowledge, our work appears to be the first one to address the multichannel rendezvous problem by using the LSH technique for mining similar objects.
  As described in the book \cite{Leskovec2020}, LSH is a technique that hashes similar items into the same bucket with high probability. Since similar items are hashed into the same buckets, the technique maximizes hash collisions.
If the available channel sets of two users are similar, then
the multichannel rendezvous problem can be viewed as a data mining problem that maximizes hash collisions.
In this paper, we first consider the synchronous setting, where the clocks of the two users are synchronized, i.e., the sym/sync/hetero/global MRP in \cite{ToN2017}.
For that setting, we propose the LSH algorithm (see Algorithm \ref{alg:LSH}) that leads to an ETTR close to the inverse of the Jaccard index between the two available channel sets, i.e., $\frac{n_1+n_2-n_{1,2}}{n_{1,2}}$. Such an ETTR is substantially smaller
than the ETTR of the random algorithm. Moreover, we propose an improved version, the LSH2 algorithm (see Algorithm \ref{alg:LSH2}), with an MTTR bounded by $N$, where $N$ is the total number of globally labeled channels.
 By conducting extensive simulations, we show that LSH2 outperforms the best algorithm in
 ETTR and MTTR for the sym/sync/hetero/global MRP.

 We then consider the asynchronous setting,  i.e., the sym/async/hetero/global MRP in \cite{ToN2017}.
 In the asynchronous setting, it is much more difficult for the two users to rendezvous as their clocks are not synchronized.
To our surprise, a direct extension of the LSH2 algorithm to the asynchronous setting, called the LSH3 algorithm in Algorithm \ref{alg:LSH3}, can still achieve a 50\% reduction of the ETTR of the random algorithm when the Jaccard index is almost 1.
 To further reduce the ETTR, we propose the LSH4 algorithm (see Algorithm \ref{alg:LSH4}) that uses LSH
 for dimensionality reduction.
 The LSH4 algorithm maps the available channel set of a user to a much smaller multiset with size $T_0$.
 We show that
 the ETTR of the LSH4 algorithm is smaller than that of the random algorithm if
 \beq{redu0000}
T_0 \le \frac{n_1 n_2}{n_1+n_2 -n_{1,2}}.
\eeq

The rest of the paper is organized as follows. In \rsec{mrp}, we briefly review the  multichannel rendezvous problem
and related works.
 In \rsec{sync}, we consider the synchronous setting, in which we propose the LSH algorithm and the LSH2 algorithm and discuss the pros and cons of these two algorithms.
In \rsec{asyn}, we  consider the asynchronous setting and propose the LSH3 algorithm and the LSH4 algorithm.  In \rsec{sim}, we provide simulation results for
comparing our algorithms with the random algorithm and the best algorithm in the literature.
We conclude the paper by discussing possible future works in \rsec{con}.

 \bsec{The multichannel rendezvous problem}{mrp}

 In this paper, we consider the multichannel rendezvous problem in  a wireless network with $N$ channels (with $N \ge 2$), indexed from $0$ to $N-1$. There are two users who would like to rendezvous on a commonly available channel by hopping over these  channels with respect to time. We assume that time is slotted (the discrete-time setting) and indexed from $t=0,1,2,\ldots$. The length of a time slot, typically in the order of 10ms, should be long enough for the two users to establish their communication link on a common channel. In the literature, the slot boundaries of these two users are commonly assumed to be aligned. If the slot boundaries of these two users  are not aligned, one can double  the size of each time slot so that the overlap of two misaligned time slots is not smaller than the original length of a time slot.

  The available channel set for user $i$, $i=1,2$, $${\bf c}_i=\{c_i(0), c_i(1), \ldots, c_i(n_i-1)\},$$
  is a subset of the $N$ channels, where $n_i=|{\bf c}_i|$ is the number of available channels to user $i$.
  We assume that there is at least one channel that is commonly available to the two users (as otherwise, it is impossible for the two users to rendezvous), i.e.,
\beq{avail1111}
{\bf c}_1 \cap {\bf c}_2 \ne \varnothing.
\eeq
Let $n_{1,2}=|{\bf c}_1 \cap {\bf c}_2 |$ be the number of common channels between these two users.

Define  the time-to-rendezvous (TTR) as the number of time slots needed for these two users to hop to a common available channel.
In this paper, we are interested in the expected time-to-rendezvous (ETTR),  and
the maximum time-to-rendezvous
 (MTTR). While ETTR is a measure of the average-case performance, MTTR is a measure of the worst-case performance.

There are several existing channel hopping algorithms in the literature for the sym/sync/hetero/global MRP \cite{SynMAC,Quorum,wjliao,ETCH2013,ToN2015,ChangGC17,CRISS2019}. Among them, SynMAC \cite{SynMAC} performs best if the channel load is not a concern. As we only consider the rendezvous problem with two users, we do not consider the load constraints \cite{ToN2015,ChangGC17} in this paper.
In SynMAC, each user simply hops to channel $c$ at time $t$ in a period of $N$ time slots. As long as there exists a common channel between the two users, SynMAC guarantees the rendezvous of the two users within $N$ time slots. Thus, its MTTR is $N$, and it achieves the optimal MTTR for the sym/sync/hetero/global MRP with $N$ channels. However, the ETTR of SynMAC could be very bad when the number of available channels for each user is much smaller than $N$. For instance, if $n_1 n_2<< N$, then the ETTR of SynMAC is $O(N)$ which is much larger than that of the random algorithm.
We also note that both the SYNC-ETCH
algorithm \cite{ETCH2013} and the criss-cross construction \cite{CRISS2019} are based on a tournament design \cite{BTD85}, and they only work in the homogeneous setting (where all the $N$ channels are available channels to the two users). To apply these algorithms in the heterogeneous setting, one could use the channel rotation trick in \cite{wjliao,ToN2015} to extend the CH sequence. However, the MTTR for such an extension is $O(N^2)$, which is much worse than $N$ in SynMAC. As such, we will use SynMAC and the random algorithm as our benchmarks for the synchronous setting.

For the sym/async/hetero/global MRP, most existing algorithms focus on the MTTR. The best-known MTTR  is $O(\log\log N)n_1 n_2$ (see, e.g., \cite{Chen14,Improved2015,Chang18}). However, their ETTRs  are not as good as that of the random algorithm. The quasi-random (QR) algorithm \cite{Quasi2018} that uses the 4B5B encoding and the two-prime clock algorithm in \cite{ToN2017} has a comparable ETTR to the random algorithm and an $O(\log N)n_1 n_2$ MTTR.
Another CH algorithm that has an $O(\log N)n_1 n_2$ MTTR is QECH \cite{QECH}. There are also periodic CH sequences with $O(N^2)$ periods
that can achieve full rendezvous diversity, see, e.g., CRSEQ \cite{CRSEQ}, DRDS \cite{DRDS13}, T-CH \cite{Matrix2015}, DSCR \cite{DSCR2016}, IDEAL-CH \cite{GAP2019}, RDSML-CH \cite{Wang2022}. These CH sequences, along with random patching, can be used for the sym/async/hetero/global MRP to achieve $O(N^2)$ MTTR. However, according to their simulation results, their ETTRs are not as good as that of the random algorithm.
One open question in the literature is whether there exist CH algorithms that have smaller ETTR than that of the random algorithm for the sym/async/hetero/global MRP.

\bsec{The synchronous setting}{sync}

In this section, we consider the sym/sync/hetero/global MRP. In such a setting,
 (i) sym: users are indistinguishable, (ii) sync:
 users' clocks {\em are synchronized} to the global clock, (iii) hetero: users might not have the same available channel sets, and (iv) global: the labels of the channels of a user are the same as the global labels of the channels.
We will use the locality-sensitive hashing (LSH) technique to generate channel hopping sequences that can outperform the random algorithm.

\bsubsec{Locality-sensitive hashing}{reviewlsh}

In this section, we briefly introduce the locality-sensitive hashing technique for mining large data sets.
For a more detailed introduction, we refer to the book \cite{Leskovec2020}.
In data mining, objects are often represented by a binary vector of $N>>1$ features, indexed from 0 to $N-1$. Given the binary vector of a specific object, we generate a set of $K$ hash functions that are uniformly distributed in $[0,N-1]$. For each hash function, it returns the index of the first nonzero element of a binary vector (in the modulo manner).
 If the $K$ hash values of another vector are the same as that of the given vector, then that vector is declared to be similar to the given vector. By doing so, searching similar objects in a large dataset can be efficiently done by only comparing the hash values.

The key rationale behind the LSH technique is that the probability that two binary vectors have the same hash value is approximately equal to the Jaccard index between these two binary vectors \cite{Leskovec2020}. The Jaccard index is a widely used similarity measure between two sets $S_1$ and $S_2$, and it is defined as
\beq{jacc1100}
J=\frac{|S_1 \cap S_2|}{|S_1 \cup S_2|}.
\eeq
Thus, the probability that the $K$ hash values of two vectors are the same is roughly $J^K$, which is very small unless the Jaccard index $J$ is close to 1.

\bsubsec{Channel hopping algorithms}{algorithm}

To map the multichannel rendezvous problem to a data mining problem, we represent
the available channel set of each user by  one-hot encoding the available channel set into an $N$-dimensional binary vector.
Since these two users are usually in the same proximity, it is reasonable to assume that the Jaccard index between the two available channel sets,
\beq{jacc1111}
J=\frac{|{\bf c}_1 \cap {\bf c}_2 |}{|{\bf c}_1 \cup {\bf c}_2 |}=\frac{n_{1,2}}{n_1+n_2-n_{1,2}},
\eeq
 is not too small. As the LSH technique for data mining, we generate a sequence of uniformly distributed  hash functions $\{U(t), t \ge 0\}$. At time $t$, each user selects the channel that is the first nonzero element from $U(t)$ in its one-hot encoded vector. This is described in Algorithm \ref{alg:LSH}.

\begin{algorithm}\caption{The LSH CH algorithm}\label{alg:LSH}
\noindent {\bf Input}:  A set of available channels ${\bf c}=\{c_0, c_1, \ldots, c_{n-1}\}$ that is a subset of the $N$ channels $\{0,1, \ldots, N-1\}$,  and a sequence of uniformly distributed pseudo-random variables $\{U(t), t \ge 0\}$ in $[0,N-1]$.

\noindent {\bf Output}: A CH sequence $\{c(t), t=0,1,\ldots \}$ with $c(t) \in {\bf c}$.


\noindent 1: Generate CH sequence $\{c(t), t =0,1,\ldots, \}$ with
 $c(t)=c_{i^*}$, where $i^*={\rm argmin}_{0 \le i \le n-1}((c_i-U(t))\;\mod\;N)$.
\end{algorithm}

\begin{figure}[ht]
\centering
\includegraphics[width=0.4\textwidth]{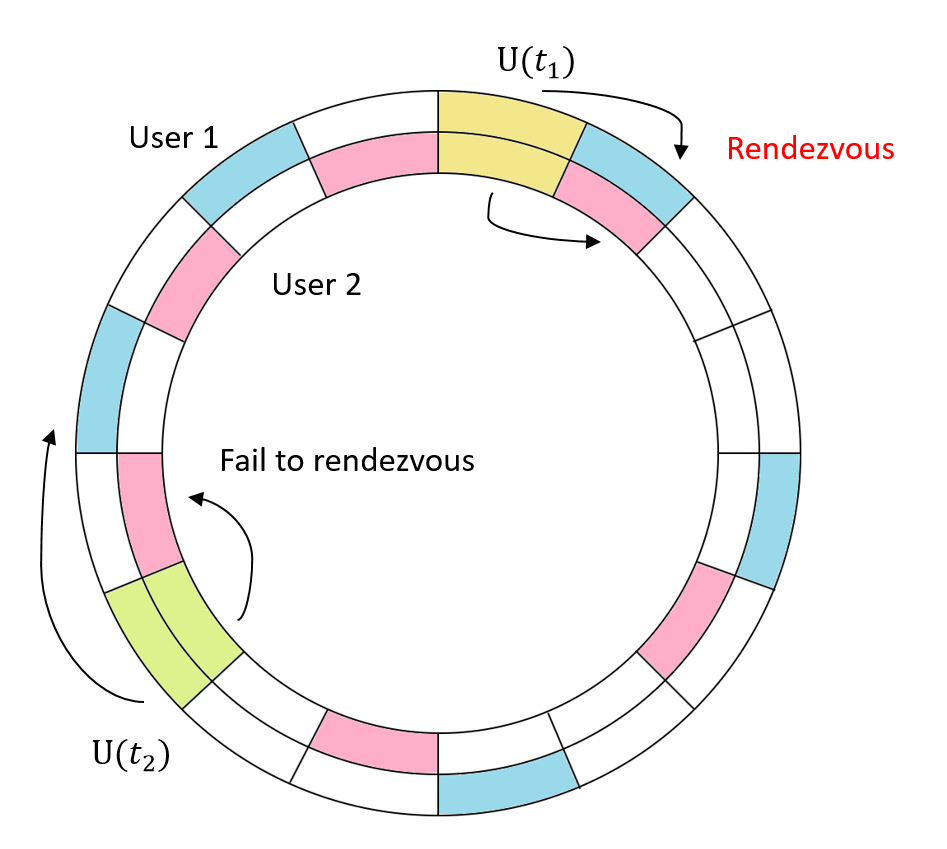}
\caption{An illustration of the LSH CH algorithm in Algorithm \ref{alg:LSH}: a successful rendezvous at time $t_1$ and a failed rendezvous at time $t_2$. }
\label{fig:ring}
\end{figure}

In \rfig{ring}, we illustrate how the LSH CH algorithm works. In this figure, the $N$ channels are ordered in a ring. The outer  ring represents the available channels of user 1 (colored in blue), and the inner
ring represents the available channels of user 2 (colored in pink). At time $t_1$, a pseudo-random variable $U(t_1)$ is selected in the ring (colored in yellow), and each user hops to the nearest available channel clockwise. In \rfig{ring}, the nearest available channel of user 1 is the same as that of user 2, and thus both users rendezvous at time $t_1$. However, at time $t_2$, another pseudo-random variable $U(t_2)$ is selected in the ring (colored in green). For $U(t_2)$, the nearest available channel of user 1 is different from that of user 2, and thus both users fail to rendezvous at time $t_2$.

The insight behind the LSH CH algorithm is that the $n_1+n_2-n_{1,2}$ common channels  partition the ring  into $n_1+n_2-n_{1,2}$ line graphs. If $U(t)$ falls in a line graph associated with a common channel, then the two users rendezvous. Otherwise, they do not. If the numbers of nodes of these line graphs are evenly distributed, then the probability that $U(t)$ falls in a line graph associated with a common channel is
$\frac{n_{1,2}}{n_1+n_2-n_{1,2}}=J$.
Since $\{U(t), t \ge 0\}$ are assumed to be independent, one would expect the ETTR of Algorithm \ref{alg:LSH} to be close to $1/J$.

\begin{figure}[ht]
\centering
\includegraphics[width=0.45\textwidth]{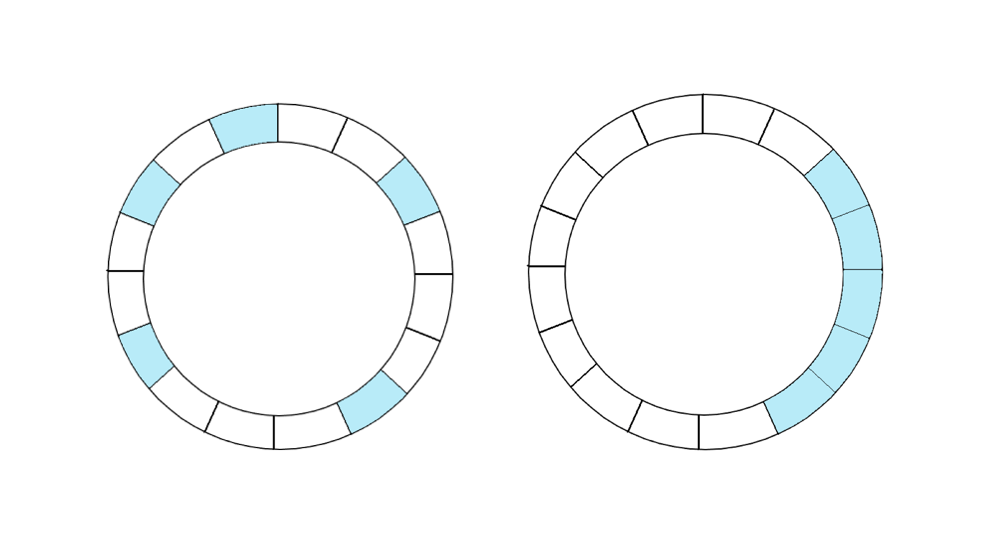}
\caption{Normally, available channels are evenly distributed (as shown in the drawing on the left). However, sometimes available channels are contiguous (as shown in the drawing on the right), and the leading channel in a contagious block has a much larger probability of being selected.
}
\label{fig:bad}
\end{figure}

However, in some of our simulations, we find that the ETTR of Algorithm \ref{alg:LSH} could sometimes be much larger than $1/J$ if the numbers of nodes of the $n_1+n_2-n_{1,2}$ line graphs are {\em not} evenly distributed. This happens when the available channels of a user are contiguous (see \rfig{bad} for an illustration).
The leading channel in a contiguous block of available channels has a much larger probability of being selected than the other available channels in the same block. This is because a non-leading available channel in the block can only be selected when $U(t)$ is exactly the same as that channel.
 To remedy this, we
add a pseudo-random permutation $\pi_1$ of the $N$ channels that reorders the $N$ channels randomly so that each available channel of a user can be hashed roughly with an equal probability.

Another drawback of the LSH CH algorithm in Algorithm \ref{alg:LSH} is that its maximum time-to-rendezvous (MTTR) is not bounded.  This is because
nothing is learned  from a failed rendezvous attempt  by using independent uniformly distributed random variables $\{U(t), t\ge 0\}$ in Algorithm \ref{alg:LSH}. In fact, when a user uses $U(t)=u$ to select a channel at time $t$ and it results in a failed rendezvous, then the user knows that the value $u$ should not be used again.
To learn from a failed rendezvous attempt,
one should use sampling without replacement. For this, we replace the sequence of uniform hash functions $\{U(t), t \ge 0\}$ by a pseudo-random permutation $\pi_2$ of the $N$ channels. This leads to the following improved algorithm, called the LSH2 CH algorithm in Algorithm \ref{alg:LSH2}.

\begin{algorithm}\caption{The LSH2 CH algorithm}\label{alg:LSH2}
\noindent {\bf Input}:  A set of available channels ${\bf c}=\{c_0, c_1, \ldots, c_{n-1}\}$ that is a subset of the $N$ channels $\{0,1, \ldots, N-1\}$,  and two pseudo-random permutations $\pi_1$ and $\pi_2$ of the $N$ channels.

\noindent {\bf Output}: A CH sequence $\{c(t), t=0,1,\ldots, N-1 \}$ with $c(t) \in {\bf c}$.


\noindent 1: Let
 $c(t)=c_{i^*}$, where
 $$i^*={\rm argmin}_{0 \le i \le n-1}((\pi_1(c_i)-\pi_2(t))\;\mod\;N).$$
\end{algorithm}

\bthe{LSH2}
Consider the sym/sync/hetero/global MRP described in \rsec{sync}.
Assume that  the two users use Algorithm \ref{alg:LSH2} to generate their CH sequences
and that the two random permutations $\pi_1$ and $\pi_2$  are independent. Then
the MTTR is bounded by $N$, and the ETTR approaches  $1/J$ when $N \to \infty$.
\ethe

\bproof
We first show that the MTTR is bounded above by $N$. Let $\pi_2^{-1}$ be the inverse permutation of $\pi_2$. 
Suppose that channel $c$ is a common channel.
Then we have from Algorithm \ref{alg:LSH2} that both users hop on channel $c$ at time $\pi_2^{-1}(\pi_1(c))$.
As such, the two users rendezvous within $N$ time slots.

Let $E_t$ be the event that the two users hop to a common channel at time $t$. We will show that
\beq{LSH6666}
\pr (E_t)= J.
\eeq
To show this, let us consider a directed ring with $N$ nodes, indexed from $0,1,\ldots, N-1$ (see \rfig{ring}). In the ring, there is a directed edge from node $i$ to node $(i+1) \mod N$.
Let ${\hat c}_0, \ldots, {\hat c}_{n_1+n_2-n_{1,2}-1}$ be the  $n_1+n_2-n_{1,2}$ common channels.
Since $\pi_1$ is a random permutation, $\pi_1({\hat c}_i)$, $i=0,1,\ldots, n_1+n_2-n_{1,2}-1$, are randomly placed one by one
(without replacement) into the ring.
Let
\beq{LSH2111}
B=\{\pi_1({\hat c}_i), i=0,1,\ldots, n_1+n_2-n_{1,2}-1\} .
\eeq
Denote by
$\sigma_i$ be the $(i+1)^{th}$ smallest number in $B$, i.e.,
$$\sigma_0 < \sigma_1 <\ldots < \sigma_{n_1+n_2-n_{1,2}-1}.$$
Now associate each node in the ring with the nearest neighbor (including itself) in $B$, and  this partitions the ring into $n_1+n_2-n_{1,2}$ line graphs, where the $\ell^{th}$ line graph contains nodes between
node $\sigma_{(\ell-1) \mod n_1+n_2-n_{1,2}}+1$ and node $\sigma_\ell$, $\ell=0, \ldots, n_1+n_2-n_{1,2}-1$.
Node $\sigma_\ell$ is called the representing node of the $\ell^{th}$ line graph, and
a line graph is called a rendezvous line graph if its representing node, say node $\sigma$, is mapped from a common channel, i.e.,  $\pi_1^{-1}(\sigma)$ is a common channel of the two users.
As there are $n_{1,2}$ common channels, there are $n_{1,2}$ rendezvous line graphs (among the $n_1+n_2-n_{1,2}$ line graphs).
Let $B_2$ be the set of nodes in the $n_{1,2}$ rendezvous line graphs.
According to Algorithm \ref{alg:LSH2}, the two users hop to a common channel at time $t$ if and only if
 $\pi_2(t)$ is a node in $B_2$.
 Since $\pi_1({\hat c}_i)$, $i=0,1,\ldots, n_1+n_2-n_{1,2}-1$, are randomly placed one by one
into the ring, it follows from the symmetry of a random permutation that the numbers of nodes in the $n_1+n_2-n_{1,2}$ line graphs are {\em identically distributed}.
 Thus,
 \beq{LSH2211}
 \pr (E_t)=\frac{\ex[|B_2|]}{N}=\frac{n_{1,2}}{n_1+n_2-n_{1,2}}=J.
 \eeq
 Note that the events $\{E_t, t =0,1,\ldots, \tau\}$ for a finite time $\tau$ are not independent.
 However,
 since $\pi_2(t)$ is a random permutation, the events $\{E_t, t =0,1,\ldots, \tau\}$ converges to  independent Bernoulli trials  when $N \to \infty$. This implies that
  the time-to-rendezvous converges to a geometric distribution with mean $1/J$ when $N \to \infty$.


\eproof

\bsec{The asynchronous setting}{asyn}

Now we extend the LSH algorithms to the asynchronous setting.
Specifically, we consider
the sym/async/hetero/global MRP.
In such a setting, (i) sym: users are indistinguishable, (ii) async:
 users' clocks {\em might not be  synchronized} to the global clock, (iii) hetero: users might not have the same available channel sets, and (iv) global: the labels of the channels of a user are the same as the global labels of the channels.

\bsubsec{Direct extension}{direct}

One simple extension is to use the LSH2 algorithm.
 As the clocks of the two users may not be synchronized, the LSH2 algorithm cannot guarantee the rendezvous of the two users. For example, if all the channels are available to the two users, i.e., $n_1=n_2=N$, and their clocks are shifted by one time slot, then the two users using the LSH2 algorithm will hop to two different channels in every time slot.
 To avoid the scenario that the two users never rendezvous, we need to add ``randomness'' into the rendezvous algorithm. This can be easily done  by using {\em sampling with replacement} in the LSH algorithm. As a combination of the LSH algorithm and the LSH2 algorithm, we propose the LSH3 algorithm in Algorithm \ref{alg:LSH3} for the sym/async/hetero/global MRP.

 \begin{algorithm}\caption{The LSH3 CH algorithm}\label{alg:LSH3}
\noindent {\bf Input}:  A set of available channels ${\bf c}=\{c_0, c_1, \ldots, c_{n-1}\}$ that is a subset of the $N$ channels $\{0,1, \ldots, N-1\}$,  a pseudo-random permutations $\pi_1$ of the $N$ channels, and a sequence of uniformly distributed pseudo-random variables $\{U(t), t \ge 0\}$ in $[0,N-1]$.

\noindent {\bf Output}: A CH sequence $\{c(t), t=0,1,\ldots \}$ with $c(t) \in {\bf c}$.


\noindent 1: Generate CH sequence $\{c(t), t =0,1,\ldots, \}$ with
 $c(t)=c_{i^*}$, where $i^*={\rm argmin}_{0 \le i \le n-1}((\pi_1(c_i)-U(t))\;\mod\;N)$.
\end{algorithm}

Assume that  the two users use Algorithm \ref{alg:LSH3} to generate their CH sequences
and that the random permutations $\pi_1$ and the sequence of uniformly distributed random variables $\{U(t), t \ge 0\}$ are independent.
Under these assumptions, we derive an approximation for the rendezvous probability of the LSH3 algorithm in the asynchronous setting.
Recall that $E_t$ is the event that the two users hop to a common channel at time $t$.
Then
 \bear{LSHa0000}
 &&\pr (E_t) \approx \frac{2n_{1,2}}{(n_1+n_2-n_{1,2})(n_1+n_2-n_{1,2}+1)}\nonumber\\
 &&+\frac{n_{1,2}}{(n_1+n_2-n_{1,2})^2}({\frac{n_1-n_{1,2}}{n_2}}+{\frac{n_2-n_{1,2}}{n_1}}) .
 \eear
 Since the events $\{E_t, t =0,1,\ldots,\}$ are independent Bernoulli trials  when $N \to \infty$,
  the time-to-rendezvous  converges to a geometric distribution with mean $1/\pr(E_t)$ when $N \to \infty$.
Thus, we can approximate the ETTR by $1/\pr (E_t)$ as in \rthe{LSH2}.
In particular, when $n_{1,2}=n_1=n_2$, the approximation from \req{LSHa0000} for the rendezvous probability of the LSH3 algorithm  is $2/n_{1,2}$, which is twice of that of the random algorithm. Thus, the LSH3 algorithm achieves a 50\% reduction of the ETTR when the Jaccard index is almost 1.
We will further verify this by simulations in \rsec{sim}.

The rest of this section is devoted to the derivation for the approximation in \req{LSHa0000}.
Without loss of generality, we assume that the clock drift between the two users is $d$ (time slots).
At time $t$, user 1 (resp. user 2) uses $U(t)$ (resp. $U(t+d)$) to generate its hopping channel at time $t$.
Consider the directed ring with $N$ nodes  in the proof of \rthe{LSH2}.
Following the argument in the proof of \rthe{LSH2}, there are $n_1+n_2-n_{1,2}$ line graphs, indexed from $0, 1,\ldots, n_1+n_2-n_{1,2}-1$.
We classify the $n_1+n_2-n_{1,2}$ line graphs into three types: (i) a rendezvous line graph: the representing node is mapped from a common channel, (ii) a type 1 line graph: the representing node is mapped from an available channel of user 1 that is not a common channel, and (iii) a type 2 line graph: the representing node is mapped from an available channel of user 2 that is not a common channel.
Since there are $n_{1,2}$ common channels,
there are $n_{1,2}$ {\em rendezvous line graphs}.
Similarly, there are $n_1-n_{1,2}$ type 1  line graphs and $n_2-n_{1,2}$ type 2 graphs.
To simplify the notations, we let $r_1= \frac{n_1-n_{1,2}}{n_1+n_2-n_{1,2}}$ and
$r_2= \frac{n_2-n_{1,2}}{n_1+n_2-n_{1,2}}$.
Note that $r_1= \frac{n_1-n_{1,2}}{n_1+n_2-n_{1,2}}$ is  the probability that a randomly selected line graph is a type 1 line graph, and $r_2= \frac{n_2-n_{1,2}}{n_1+n_2-n_{1,2}}$ is  the probability that a randomly selected line graph is a type 2 line graph.

Let $L_\ell$, $\ell=0,1, \ldots, n_1+n_2-n_{1,2}-1$, be the number of nodes in the $\ell^{th}$ line graph. As argued in the proof of \rthe{LSH2}, it follows from the symmetry of a random permutation that the numbers of nodes in the $n_1+n_2-n_{1,2}$ line graphs are {\em identically distributed}. Moreover,
 when $N>>1$, we know from order statistic (see, e.g., the book \cite{Probbook}) that the distribution of the number of nodes of a line graph normalized to $N$, i.e., $L_\ell/N$, can be approximated by the beta distribution with the two parameters $1$ and $n_1+n_2-n_{1,2}-1$, i.e., $\mbox{Beta}(1,n_1+n_2-n_{1,2}-1)$. Recall that the first moment
 and the second moment of $\mbox{Beta}(\alpha,\beta)$ are $\frac{\alpha}{\alpha+\beta}$ and
 $\frac{\alpha(\alpha+1)}{(\alpha+\beta)(\alpha+\beta+1)}$, respectively.

Without loss of generality, suppose that the $\ell^{th}$ line graph is a rendezvous line graph with its representing node mapped from a common channel $c$. Let $\ell_1^*$ be the index of the line graph such that
 the $\ell_1^*$ line graph is {\em not} a type 1 line graph and all the line graphs  $\ell_1^*+1, \ldots, \ell-1$,
 are type 1 line graphs. Also, let $\ell_2^*$ be the index of the line graph such that
 the $\ell_2^*$ line graph is {\em not} a type 2 line graph and all the line graphs  $\ell_2^*+1, \ldots, \ell-1$,
 are type 2 line graphs.
There are three cases that the two users hop to the common channel $c$ at time $t$.

\noindent {\em Case 1}. both  $U(t)$ and $U(t+d)$ are in the $\ell^{th}$ line graph:

 Since $U(t)$ and $U(t+d)$ are assumed to be independent given the random permutation $\pi_1$, the probability that $U(t)$ and $U(t+d)$ are in the $\ell^{th}$ line graph is $\frac{L_{\ell}}{N}\frac{L_{\ell}}{N}$.
 As the distribution of $L_\ell$ can be approximated by $\mbox{Beta}(1,n_1+n_2-n_{1,2}-1)$, the probability that
 the two users hop to the common channel $c$ at time $t$ for this case is
 \beq{LSHa4444}
 \ex[\frac{L_{\ell}}{N}\frac{L_{\ell}}{N}]= \frac{2}{(n_1+n_2-n_{1,2})(n_1+n_2-n_{1,2}+1)}.
 \eeq

\noindent {\em Case 2}. $U(t)$ is the $\ell^{th}$ line graph
and $U(t+d)$ is in the $\ell_1^{th}$ line graph with $\ell_1^* < \ell_1 < \ell$:

From the definition of $\ell_1^*$, we know that
the line graphs $\ell_1^*+1, \ldots, \ell-1$ are all type 1 line graphs. Thus, user 2 will also hop to channel $c$ at time $t$.
The probability for this case is
$$\frac{L_{\ell}}{N}\frac{\sum_{\ell_1=\ell_1^*+1}^{\ell-1}L_{\ell_1}}{N}.$$
When $N>>1$, we can approximate $L_\ell$ and $L_{\ell_1}$ as two uncorrelated random variables with the distribution
$\mbox{Beta}(1,n_1+n_2-n_{1,2}-1)$. Moreover, as the probability that a line graph is a type 1 line graph is
$r_1$,
the number of line graphs from $\ell_1^*+1$ to $\ell-1$ can be approximated by the number of successive independent Bernoulli trials until the ``first'' non type 1 line graph, and this leads to on average $\frac{r_1}{1-r_1}$ line graphs. Thus, the probability that
 the two users hop to the common channel $c$ at time $t$ for this case is  roughly
\beq{LSHa5555}
 \ex[\frac{L_{\ell}}{N}\frac{\sum_{\ell_1=\ell_1^*+1}^{\ell-1}L_{\ell_1}}{N}]= \frac{1}{(n_1+n_2-n_{1,2})^2}{\frac{r_1}{1-r_1}}.
 \eeq

\noindent {\em Case 3}. $U(t+d)$ is the $\ell^{th}$ line graph
and $U(t)$ is in the $\ell_2^{th}$ line graph with $\ell_2^* < \ell_2 < \ell$:

The argument for this case is similar to Case 2.
The probability that
 the two users hop to the common channel $c$ at time $t$ for this case is  roughly
\beq{LSHa6666}
 \ex[\frac{L_{\ell}}{N}\frac{\sum_{\ell_2=\ell_2^*+1}^{\ell-1}L_{\ell_2}}{N}]= \frac{1}{(n_1+n_2-n_{1,2})^2}{\frac{r_2}{1-r_2}}.
 \eeq

Summing over these three cases, the probability that
 the two users hop to the common channel $c$ at time $t$ is roughly
   \bear{LSHa7777}
 &&\frac{2}{(n_1+n_2-n_{1,2})(n_1+n_2-n_{1,2}+1)}\nonumber\\
 &&+  \frac{1}{(n_1+n_2-n_{1,2})^2}({\frac{r_1}{1-r_1}}+{\frac{r_2}{1-r_2}}).
 \eear
 As there are $n_{1,2}$ rendezvous line graphs, the probability that
 the two users hop to the common channel $c$ at time $t$ is roughly
  \bear{LSHa8888}
 &&\pr (E_t) \approx \frac{2n_{1,2}}{(n_1+n_2-n_{1,2})(n_1+n_2-n_{1,2}+1)}\nonumber\\
 &&+\frac{n_{1,2}}{(n_1+n_2-n_{1,2})^2}({\frac{r_1}{1-r_1}}+{\frac{r_2}{1-r_2}}) .
 \eear

\bsubsec{Dimensionality reduction}{reduction}

Our second approach is to use LSH for dimensionality reduction.
The idea is to map the available channel set of a user to a much smaller multiset (that might have duplicated elements).
For this, we use the LSH2 algorithm to generate a CH sequence with the length $T_0$. Let $\tilde {\bf c}_i=[\tilde c_i(0), \tilde c_i(1),\ldots, \tilde c_i(T_0-1)]$ be the multiset from the CH sequence
by user $i$, $i=1$ and 2. The size of the multiset $T_0$ is fixed, and it should be much smaller than $n_1$ and $n_2$ for dimensionality reduction.
Since the multisets are generated by the LSH2 algorithm, we have from \rthe{LSH2} that there are, on average, $J T_0$ times such that $\tilde c_1(t)=\tilde c_2(t)$, $t=0,1,2, \ldots, T_0-1$. Thus, the size of the intersection of the two multisets is roughly $J T_0$.
In the asynchronous setting,
suppose that at each time slot each user randomly selects an element from its multiset and hops to that channel. Then at time slot $t$, the two users will rendezvous with probability roughly $J T_0/(T_0^2)$, and the ETTR for such an algorithm  is roughly $T_0/J$, which is only $T_0$ times larger than that in the synchronous setting. Moreover, it is smaller than
the ETTR of the random algorithm if
\beq{redu1111}
T_0 \le \frac{n_1 n_2}{n_1+n_2 -n_{1,2}}.
\eeq

One problem with the above argument is that there is a nonzero probability that the intersection of the two multisets is empty. If this happens, the two users never rendezvous. One quick fix is to use a mixed strategy. With probability $p$ (resp. $(1-p)$), a user selects a channel from its multiset (resp. available channel set). This leads to the LSH4 algorithm in Algorithm \ref{alg:LSH4}.

\begin{algorithm}\caption{The LSH4 CH algorithm}\label{alg:LSH4}
\noindent {\bf Input}:  A set of available channels ${\bf c}=\{c_0, c_1, \ldots, c_{n-1}\}$ that is a subset of the $N$ channels $\{0,1, \ldots, N-1\}$, two pseudo-random permutations $\pi_1$ and $\pi_2$ of the $N$ channels, and two parameters $T_0$ and $p$.

\noindent {\bf Output}: A CH sequence $\{c(t), t=0,1,\ldots, N-1 \}$ with $c(t) \in {\bf c}$.


\noindent 1: Generate the multiset $\tilde {\bf c}=[\tilde c(0), \tilde c(1), \ldots, \tilde c(T_0-1)]$ by  letting
 $\tilde c(t)=c_{i^*}$, where
 $$i^*={\rm argmin}_{0 \le i \le n-1}((\pi_1(c_i)-\pi_2(t))\;\mod\;N).$$

\noindent 2:  With probability $p$ (resp. $(1-p)$), randomly select an element $c$ from $\tilde {\bf c}$
(resp. ${\bf c}$) and let $c(t)=c$.

\end{algorithm}

One can approximate the ETTR of the LSH4 algorithm by considering the four cases: (i) with probability $(1-p)^2$, both users select their channels from the available channel sets, (ii) with probability $p(1-p)$, user 1 selects its channel from its multiset, and user 2 selects its channel from its available channel set, (iii) with probability $p(1-p)$,
user 2 selects its channel from its multiset, and user 1 selects its channel from its available channel set, and (iv)
with probability $p^2$, both users select their channels from the multisets. As argued before, the probability for a successful rendezvous in the fourth case is roughly $J/T_0$. Also, the probability of a successful rendezvous in the first case is simply the rendezvous probability of the random algorithm, i.e., $\frac{n_{1,2}}{n_1 n_2}$. For the second case, we have to average over all the random permutations $\pi_1$ for user 1, and the probability that user 1 selects a channel $c$ (after the averaging) is the same as that of the random algorithm. Hence the probability of a successful rendezvous in the second case is also the same as that of the random algorithm, i.e., $\frac{n_{1,2}}{n_1 n_2}$.
The argument for the third case is the same as that for the second case. Averaging over these four cases,
the probability of a successful rendezvous in a time slot is
$$(1-p^2) \frac{n_{1,2}}{n_1 n_2}+ p^2 \frac{J}{T_0}.$$
As such, the ETTR of the LSH4 algorithm is approximately
\beq{LSH41111}
 \frac{1}{(1-p^2) \frac{n_{1,2}}{n_1 n_2}+ p^2 \frac{J}{T_0}}.
 \eeq

\bsec{Simulations}{sim}

\bsubsec{Numerical results in the synchronous setting}{numsyn}

In this section, we compare our LSH CH algorithms in Algorithm \ref{alg:LSH} and Algorithm \ref{alg:LSH2} with the
random algorithm and the SynMAC algorithm \cite{SynMAC}. For SynMAC, if $t$ is not in the available channel set of a user, we simply replace it by randomly selecting a channel in the available channel set (random patching).

For our experiments, we first specify the number of channels $N$, the number of common channels $n_{1,2}$, the number of available channels $n_1$ for user 1, and the number of available channels $n_2$ for user 2.
We then  randomly select $n_{1,2}$ channels from $\{0,1,\ldots, N-1\}$ as the common channels of the two users, i.e., ${\bf c}_1\cap {\bf c}_2$. Remove these channels from $\{0,1,\ldots, N-1\}$ to form a set of channels $C$.
Randomly select two disjoint channel sets $A_1$ and $A_2$ from $C$, where $A_1$ has $n_1-n_{1,2}$ channels and
$A_2$ has $n_2-n_{1,2}$ channels. Add the channels in $A_1$ (resp. $A_2$) to ${\bf c_1}\cap {\bf c_2}$ to form the available channel set ${\bf c}_1$ for user 1 (resp. ${\bf c}_2$ for user 2).
By doing so, the Jaccard index $J$ between the two available channel sets is
$\frac{n_{1,2}}{n_1+n_2-n_{1,2}}$
For each experiment, we conduct the simulation for 10,000 time slots and record the rendezvous time slots.
The ETTR  of each experiment is the average of the TTRs in the 10,000 time slots, and the MTTR of
each experiment is the maximum of the TTRs in the 10,000 time slots.
We conduct
10,000 experiments and obtain our estimates for ETTR and MTTR by averaging over the ETTRs and MTTRs of these 10,000 experiments.

\begin{figure}[ht]
\centering
\includegraphics[width=0.4\textwidth]{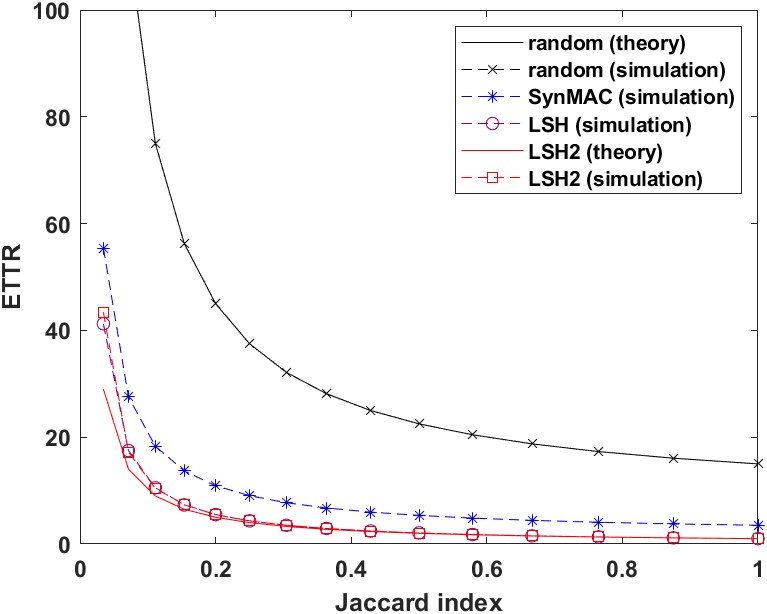}
\caption{Comparisons of the ETTRs in the synchronous setting with $N$=64, $n_{1}=n_{2}=15$.}
\label{fig:compareETTR}
\end{figure}

\begin{figure}[ht]
\centering
\includegraphics[width=0.35\textwidth]{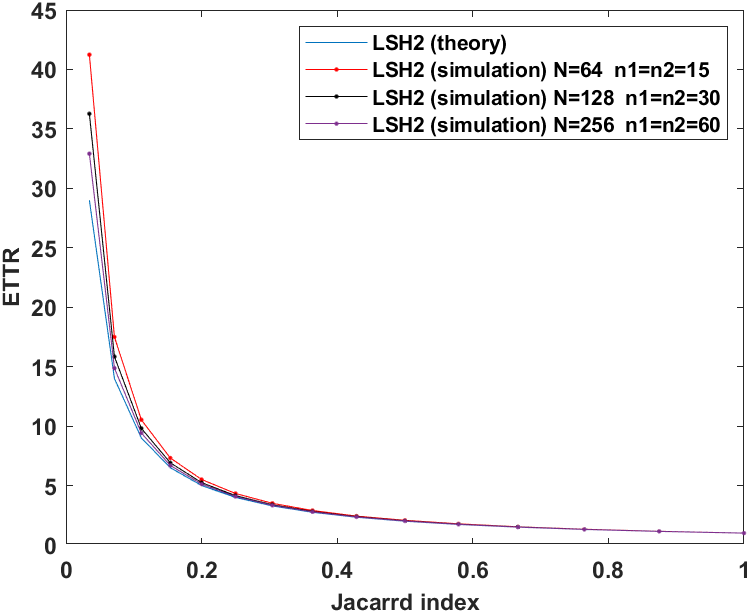}
\caption{The ETTRs of the LSH2 algorithm for $N=64, 128$, and 256.}
\label{fig:limit}
\end{figure}

In \rfig{compareETTR}, we show the ETTRs  as a function of the Jaccard index when $N=64$, $n_1=n_2=15$.
As $J=n_{1,2}/(n_1+n_2-n_{1,2})$, this figure can also be interpreted as a function of the number of common channels.
As shown in \rfig{compareETTR}, increasing the Jaccard index decreases the ETTR for all these three algorithms.
Moreover, we find that LSH2 has a great improvement compared with the random algorithm and SynMAC. Specifically,
the LSH2 algorithm achieves 43\% improvement compared with SynMAC. As expected, the experimental results of the random algorithm are very close to the theoretical values, i.e., $\frac{n_1n_2}{n_{1,2}}$. While the ETTR of LSH2 is slightly higher than the derived theoretical value $1/J$ in \rthe{LSH2}. This is because the number of channels $N$ is only 64. We also conduct simulations for other choices of channel numbers, including $N=128, n_{1}=n_{2}=30$ and $N=256, n_{1}=n_{2}=60$.
When the number of channels increases from 64 to 256, the simulation results confirm that the ETTR approaches to the theoretical value $1/J$
 as shown in  \rfig{limit}.


\begin{figure}[ht]
	\centering
	\includegraphics[width=0.4\textwidth]{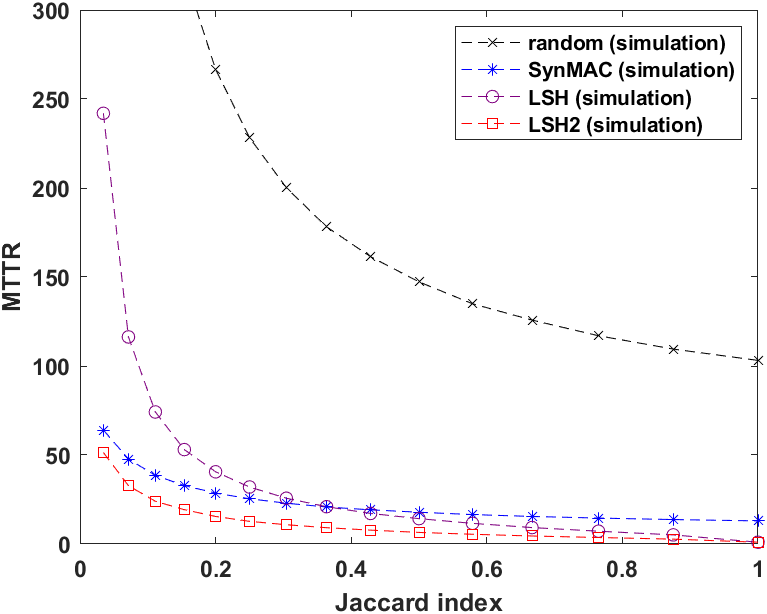}
	\caption{Comparisons of the (measured) MTTRs in the synchronous setting with $N$=64, $n_{1}=n_{2}=15$.}
	\label{fig:MTTR}
\end{figure}

In \rfig{MTTR}, we compare the
MTTRs of the LSH algorithm and the LSH2 algorithm with the random algorithm and SynMAC.
As shown in \rfig{MTTR}, the (measured) MTTR of the random algorithm is much larger than the other algorithms as it is not bounded in theory.
We also find that the MTTR of LSH is slightly larger than SynMAC when the Jaccard index is small. It is because LSH uses sampling with replacement, and it is more likely to repeatedly select the same value of $U(t)$ when the Jaccard index is small.
By using sampling without replacement in LHS2, the MTTR of LSH2 is greatly improved, and it outperforms SynMAC for all the selections of the Jaccard index.
To conclude, our numerical results show that there is roughly a 41\% to 49\% reduction in the ETTR and MTTR for LSH2
 when compared with SynMAC.

\bsubsec{Numerical results in the asynchronous setting}{numasyn}

\begin{figure}[ht]
	\centering
	\includegraphics[width=0.4\textwidth]{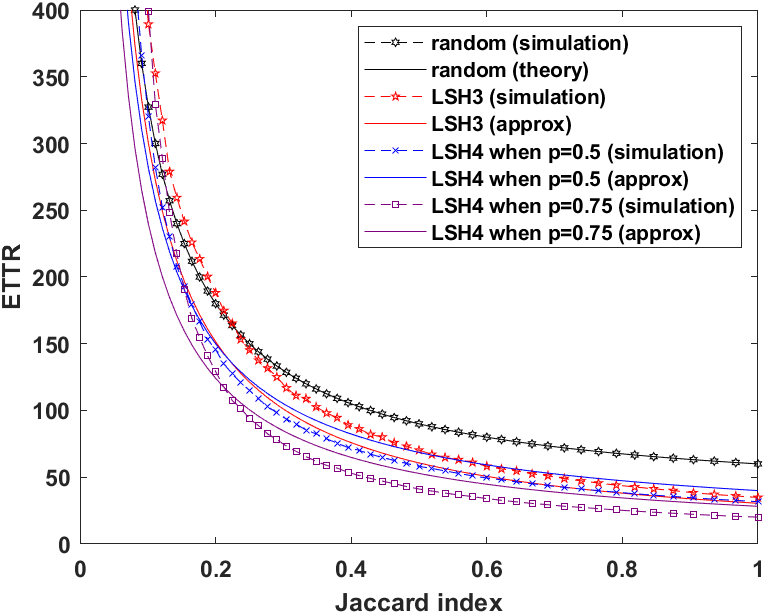}
	\caption{Comparisons of the ETTRs in the asynchronous setting with  $N$=256, $n_{1}=n_{2}=60$.}
	\label{fig:LSH3ETTR}
\end{figure}

\begin{figure}[ht]
	\centering
	\includegraphics[width=0.4\textwidth]{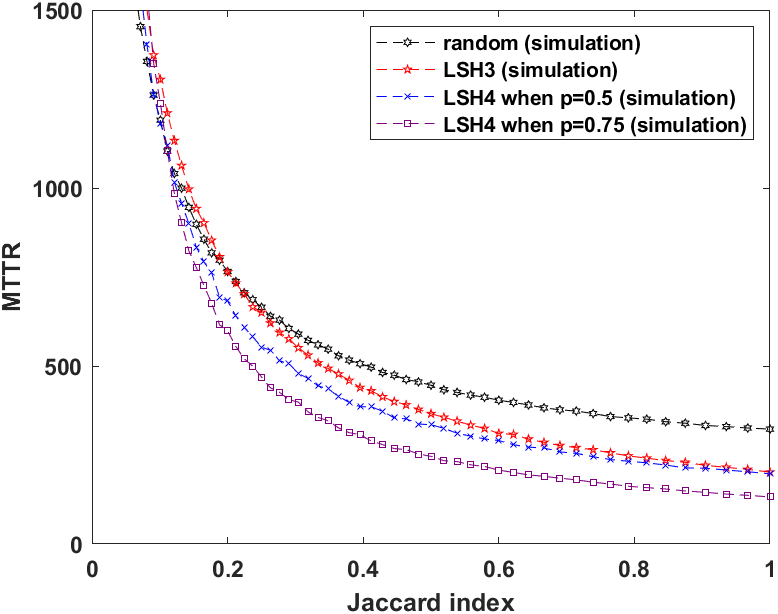}
	\caption{Comparisons of the MTTRs in the asynchronous setting with  $N$=256, $n_{1}=n_{2}=60$.}
	\label{fig:LSH3MTTR}
\end{figure}

In \rfig{LSH3ETTR} and \rfig{LSH3MTTR}, we compare the ETTRs and the MTTRs in the asynchronous setting with  $N$=256, $n_{1}=n_{2}=60$. For the LSH4 algorithm, we consider two parameter settings: (i) $T_0=20$ and $p=0.5$,
and (ii) $T_0=20$ and $p=0.75$.
 As shown in \rfig{LSH3ETTR},
the LSH3 algorithm outperforms the random algorithm when the Jaccard index is larger than 0.3.
In particular, when the Jaccard index is almost 1, the ETTR of the LSH3 algorithm is only one-half of that of the random algorithm.
 On the other hand, for the two choices of $p$, the LSH4 algorithm is significantly better than the random algorithm for the whole range of the Jaccard index in this experiment. We note that there are some discrepancies
 between the approximations for the LSH3 algorithm (from \req{LSHa0000}) and the LSH4 algorithm (from
 \req{LSH41111}) and the simulation results.
 This is because the events $E_t$'s are not independent. Also, as shown in \rfig{LSH3MTTR}, the MTTRs of the LSH3 algorithm and the LSH4 algorithm are smaller than that of the random algorithm.

\bsec{Conclusion}{con}

In this paper, we used the locality-sensitive hashing technique for the multichannel rendezvous problem.
For the synchronous setting, we proposed two algorithms: LSH in Algorithm \ref{alg:LSH} and LSH2 in Algorithm \ref{alg:LSH2}.
We proved that the MTTR of LSH2 is bounded by $N$, and the ETTR of LSH2 converges to $1/J$ when $N \to \infty$.
Our simulation results show that LSH2 outperforms not only the random algorithm but also the best algorithm in the literature. For the asynchronous setting, we also proposed two algorithms: LSH3 in Algorithm \ref{alg:LSH3} and LSH4 in Algorithm \ref{alg:LSH4}. LSH3 is a direct extension of  LSH2, and it still outperforms the random algorithm when the Jaccard index is large. On the other hand, LSH4 uses the idea of dimensionality reduction to speed up the rendezvous process, and it outperforms the random algorithm in the whole range of the Jaccard index in our experiment.
However, like the random algorithm, the MTTR of the LSH4 algorithm is not bounded.
One possible extension of our work is to embed  the multiset into a
periodic CH sequence with full rendezvous diversity. By doing so, the MTTR can still be bounded with high probability.

\end{document}